\documentstyle[twoside,epsfig]{article}
\begin{document}
\setcounter{page}{1}
\begin{flushright}
MPP-2009-6
\end{flushright}
\bigskip
\begin{center}{ \Large 
{\bf Early Top Physics with ATLAS}}
\end{center}
\begin{center}
J.~Schieck \\Max-Planck-Institut f\"ur Physik,\\
F\"ohringer Ring 6 \\ 80805 M\"unchen, Germany.\\
On behalf of the ATLAS Collaboration
\end{center}
\bigskip
\abstract{The ATLAS detector is one of the two multi-purpose experiments 
located at the Large Hadron Collider (LHC) at CERN and is expected to collect 
first collision data in summer 2009. Due to the large top-quark production cross-section 
the LHC will function as a top-quark factory allowing to measure top-quark
properties even at initial luminosities. 
 We present some recently-performed studies, focussing on measurements of
the top pair and single top production cross-sections with the first  $\mathrm{fb}^{-1}$
of data.
The potential for the measurement of other top-quark 
properties like the mass will be also briefly discussed.}
\subsection*{Introduction}
The top-quark was discovered at Fermilab in 1995 and its discovery completed
the experimental verification of the three generation structure of the Standard Model. The top-quark
decays rapidly without forming a hadron and almost exclusively into
a W-boson and a b-quark. In about $1/3$ of all cases the W-boson decays 
in a highly-energetic lepton and the corresponding neutrino, leading
to a clear experimental signature in the detector.  \\
The center-of-mass energy (CME) for inital data-taking  is
currently under discussion, but will 
be most likely below 14~TeV. All results presented here are based on simulated 
events using the ATLAS detector
at a CME of 14~TeV. 
During the first year, one can expect to collect $\approx 100 \mathrm{pb}^{-1}$ with
initial luminosity of $\approx 10^{31} \mathrm{cm^{-2}s^{-1}}$. As the luminosity increases, one may
reach 1 $\mathrm{fb}^{-1}$ of data.
A more detailed discussion of the results presented  
in this paper can be found in~\cite{ATLASCSC}.
\subsection*{Top Quark Pair Production Cross-Section}
\begin{figure}[h]
\centering
\begin{tabular}{cc}
\epsfig{file=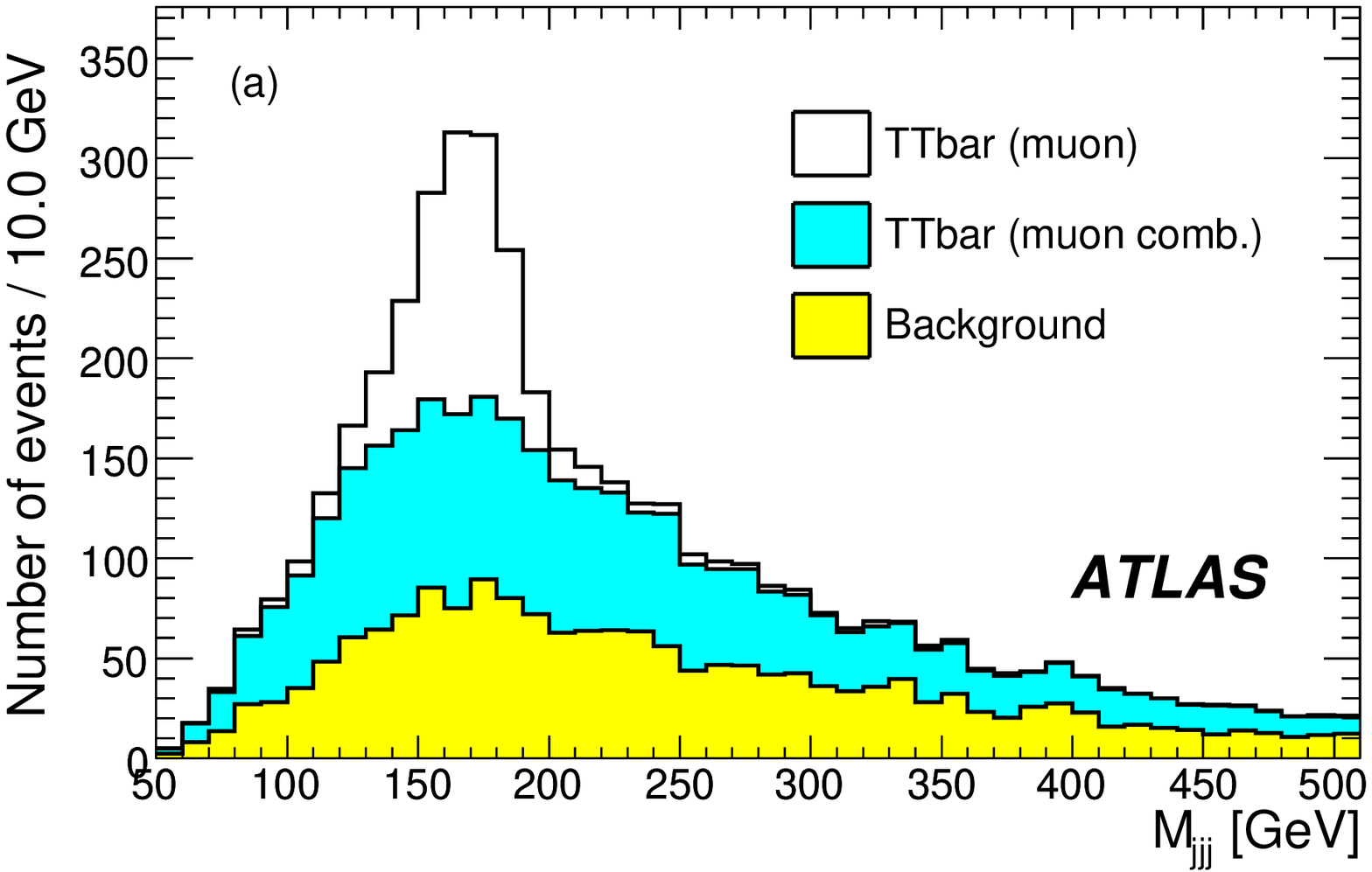,width=0.5\linewidth,angle=0}
&
\epsfig{file=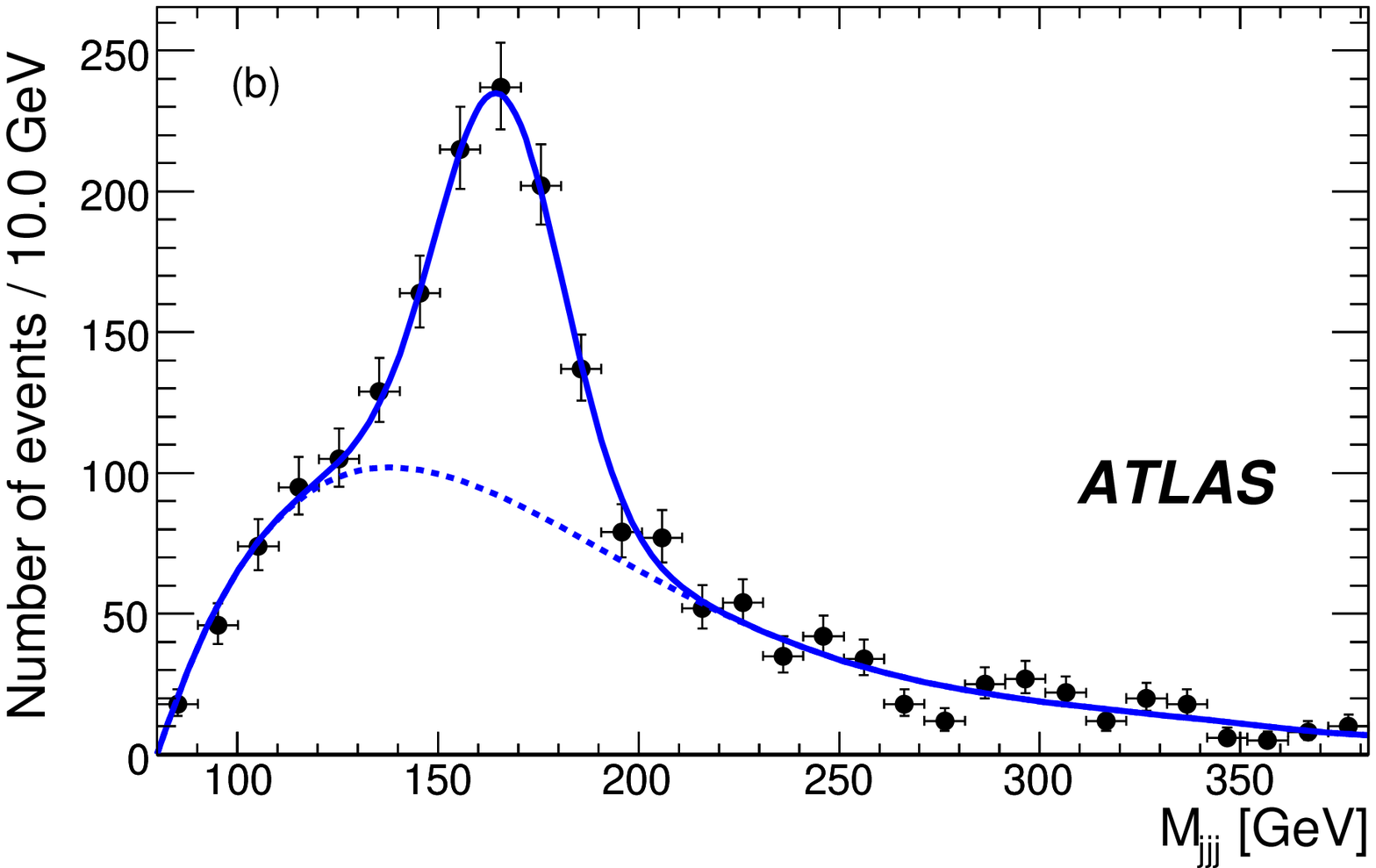,width=0.5\linewidth,angle=0} \\
\end{tabular}
\caption{(a) Expected distribution of the three-jet invariant mass after the
standard selection. 
(b) Fit to the three-jet invariant mass top-signal. Contributions from background
events are suppressed by applying an additional W-boson constraint.
The background
obtained from a Chebychev polynomial fit is indicated by a dotted line.
The gaussian fit to the signal events is indicated by the full line.}
\label{PairCrossSection}
\end{figure}

At the LHC the production of top-quark pair events is dominated by
gluon fusion processes and only about $10\%$ of all top-quark pairs
are produced via quark-antiquark annihilation.
The expected top-quark pair production cross-section $\sigma_{\mathrm pair}$ at 
a CME of 14~TeV 
is about 900 pb and is about a factor two smaller at a CME of 10~TeV. 
The ``golden channel'' to identify the production of top-quark pair
events is through the identification of the decay of one W-boson 
into a lepton and a neutrino, with the lepton being an electron or a muon, while
the other W-boson decays hadronically. 
In this final state one top-quark decays in a three jet final state and
the other decays into a jet and a high  $\mathrm{p_{T}}$ lepton.
The main background are events with a W-boson and several jets.
First measurements of the top-quark pair production cross-section
will be performed without identifying the b-quark from the
top-quark decay, leading to an increased contribution from 
background processes but decreased systematic uncertainties
from the b-quark identification.
Events are identified by requiring a high $\mathrm{p_{T}}$ lepton,
at least four high $\mathrm{p_{T}}$ jets and missing energy.
Using these cuts about $20\%$ of the $t\bar t$-events with one 
W-boson decaying hadronically and the other one decaying in a 
lepton and a neutrino are selected. 
The combination of the three out of four jets with the
largest transverse momentum is identified as the hadronically decaying
top-quark candidate. 
Figure~\ref{PairCrossSection} (a) shows the three-jet invariant
mass distribution for the muon channel after the standard selection.
The signal is indicated with the white area, while the dark shaded
 area shows the combinatorial background,
where the wrong jet combination has been selected,
and the light shaded area represents the background contribution
from events with a W-boson and several jets.
To further decrease the number of events originating 
from background events we require that at least one di-jet pair combination
of the hadronically decaying top quark is within 10~GeV of the 
mass of the W-boson. This additional cut reduces the 
selection efficiency to about $10\%$.\\
To determine $\sigma_{\mathrm pair}$
two methods are investigated. In a first method a likelihood
fit is applied to the three-jet invariant mass distribution to estimate the number
of selected  $t\bar t$-events, while the second is a  cut-and-count method.
Figure~\ref{PairCrossSection} (b) shows the result of the likelihood fit to
the three-jet invariant mass in the muon decay channel.
With 100 $\mathrm{pb}^{-1}$ of data and using the likelihood fit method we expect to reach 
$\Delta\sigma_{\mathrm pair} / \sigma_{\mathrm pair} = (7(stat.)\pm 15 (sys) \pm 3(pdf) \pm 5 (lumi))\%$
\footnote{Pdf stands for the uncertainty originating from the parton density function and lumi for 
the uncertainty from the total integrated luminosity.}. For 
the counting method $\Delta\sigma_{\mathrm pair} / \sigma_{\mathrm pair} = (3(stat.)\pm 16 (sys) \pm 3(pdf) \pm 5 (lumi))\%$
is expected. Due to the high production cross-section the accuracy of the 
analysis will be limited by systematic uncertainties. The systematic uncertainties
are dominated by the evaluation of
the W+jets background and the jet-energy scale in the case of the counting
method and by the shape of the fit function in the likelihood method.
\subsection*{Single Top Quark  Production Cross-Section}
\begin{figure}[h]
\centering
\begin{tabular}{cc}
\epsfig{file=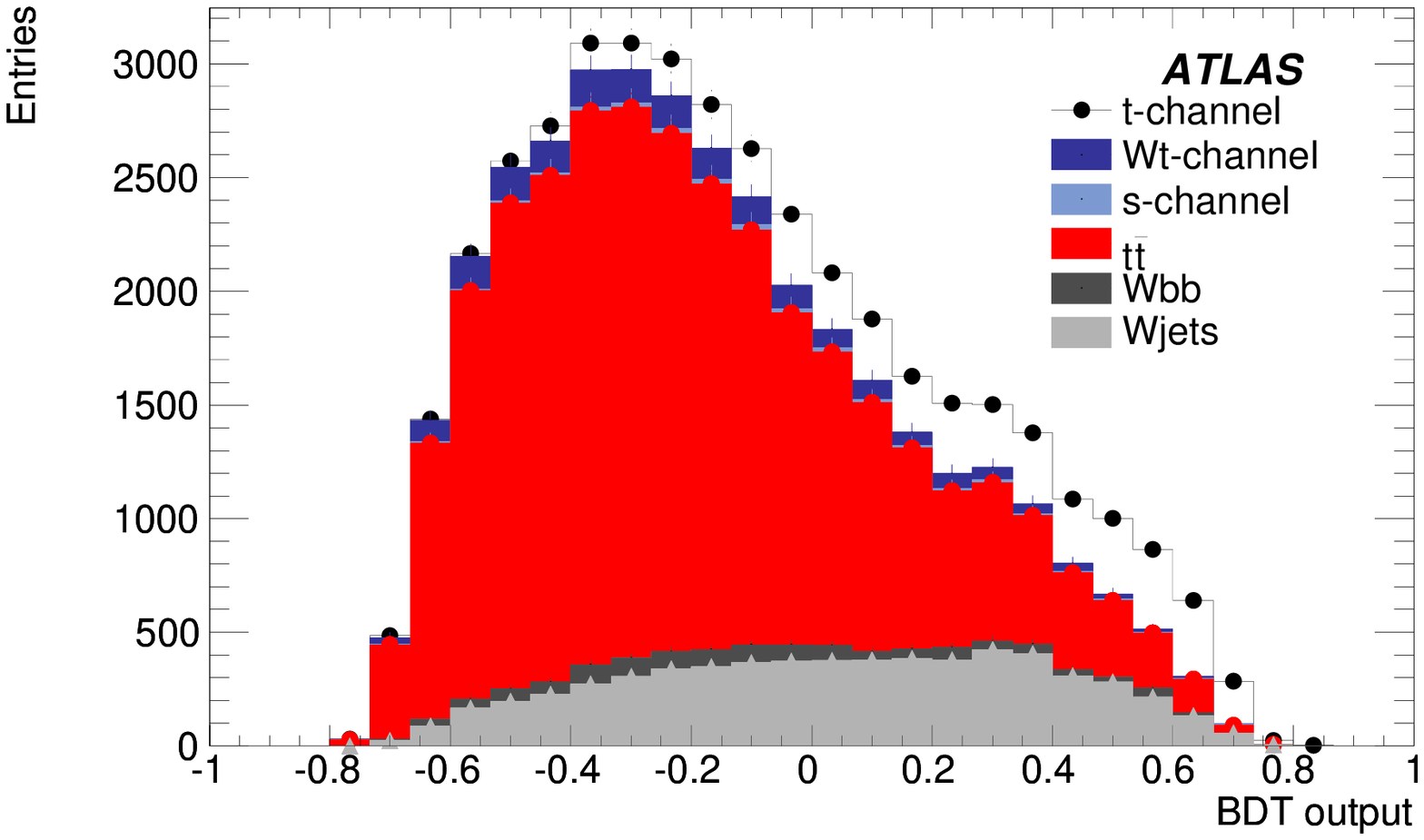,width=0.5\linewidth,clip=0} &
\begin{tabular}[b]{r|r|r}
production & stat.&  syst. \\ 
channel & uncertainty & uncertainty \\
\hline
$t$-channel  & $5\%$ ($2\%$) & $45\%$  ($22\%$) \\ \hline
$t$-channel & $6\%$  ($2\%$)& $22\%$  ($10\%$)\\
$Wt$-channel & $21\%$  ($7\%$)& $48\%$  ($19\%$)\\
$s$-channel & $64\%$  ($20\%$) & $94\%$ ($48\%$) \\
\hline 
\end{tabular} 
\end{tabular}
\caption{The Figure shows the BDT output for single-top events produced 
in the t-channel. The different signal and background contributions are indicated. 
Events with an BDT-value larger than 0.6 are used for the measurement of the cross-section.
The table on the right summarizes the expected statistical
and systematical uncertainty for the three
single-top cross-section measurements  evaluated for an integrated
luminosity of 1 (10) $\mathrm{fb}^{-1}$ of data. The first line corresponds to a cut and count method, while
the three others correspond to a BDT method.}
\label{SingleCrossSection}
\end{figure}

The production cross-section for single-top events is significantly
smaller than the top-quark pair production cross-section. There are three
possible production channels which can lead to a single top-quark
in the final state: the t-channel with an expected cross-section of
about 240 pb, the tW-channel with an expected cross-section of about 
60 pb and the s-channel with an expected cross-section of about 10 pb, all at 14~TeV. 
Measurements of
the single-top cross-section are sensitive to the production of new particles and
flavour changing neutral currents (FCNC) and allows the determination  
of the CKM-matrix element $\mathrm{V_{tb}}$.
The background contribution is higher than for
the top-quark pair production and originates predominantly from
top-quark pair events, QCD and W+jet events. \\
For the event selection  only top-quark decays with the W-boson 
decaying into an electron or a muon are considered. Events  with at least
one highly-energetic isolated lepton, two high $p_{T}$ jets, where 
one jet is identified as a b-quark jet, and missing transverse energy in the
transverse plane are considered. With a purely cut based analysis 
it is very difficult to reduce the background sufficiently to allow 
a measurement of the cross-section. For this reason multivariante
analysis methods, like the boosted decision tree (BDT) method, are used
to  determine the production cross-section. Figure~\ref{SingleCrossSection}
shows the BDT-output variable for the selection of t-channel events, 
separated for signal and background events.
For a luminosity of 1~$\mathrm{fb}^{-1}$ about 500 signal 
events are selected using a BDT-cut of 0.6 with a signal to background ratio of 1.3.
The measurement of the single top-quark production in the 
Wt- and the s-channel is considerably more difficult, 
with signal to background ratios of less than one. In order
to firmly establish the signal at least 1-10~$\mathrm{fb}^{-1}$ of data
are required. 
The expected statistical and systematic uncertainties on the 
measuremnets for the three different single top-quark production
cross-sections are shown in Figure~\ref{SingleCrossSection}.
\subsection*{Further Top Analysis}
With a data sample of 1~$\mathrm{fb}^{-1}$ it is also possible to determine precisely
the mass of the top-quark.
As for the measurement of the pair production cross-section,
a very pure sample of reconstructed top hadronic decays is obtained in a
sample of $t\bar t$-pairs where the other top decays leptonically. 
Due to the large production cross-section the statistical uncertainty is negligible
and the overall uncertainty is
dominated by the knowledge of the jet energy scale, leading to an uncertainty on the mass of 1 to 3.5~GeV
assuming a jet energy scale uncertainty of 1 to $5\%$.
A precision of the order of $30-50\%$ on the parameters describing spin
correlation  is expected with 1~$\mathrm{fb}^{-1}$. The limit on the branching fraction 
is $\approx~10^{-2}$ for the $\mathrm{t\to qg}$ decay and $10^{-3}$ for $\mathrm{t \to q\gamma/Z}$ decays.
\subsection*{Conclusion and Outlook}
The ATLAS experiment is expected to collect in the course of next year the
first data from proton-proton  collisions. Due to the large number of 
produced top-quark pairs the top-quark pair production 
cross-section $\sigma_{\mathrm pair}$ can be already measured with an accuracy 
of $\Delta\sigma_{\mathrm pair}/\sigma_{\mathrm pair} \approx 18\%$ using 100~$\mathrm{pb}^{-1}$.
A precise determination of single top quark cross-sections can be achieved
with a few $\mathrm{fb}^{-1}$ in the t-channel and Wt-channel, while for the s-channel
higher statistics will be required.

\end{document}